\documentclass[conference]{IEEEtran}
\IEEEoverridecommandlockouts
\usepackage{cite}
\usepackage{amsmath,amssymb,amsfonts}
\usepackage{algorithmic}
\usepackage{graphicx}
\usepackage{textcomp}
\usepackage{xcolor}
\usepackage{svg}
\usepackage{multibib}
\usepackage{float}
\usepackage{url}
\usepackage{hyperref}

\def\BibTeX{{\rm B\kern-.05em{\sc i\kern-.025em b}\kern-.08em
    T\kern-.1667em\lower.7ex\hbox{E}\kern-.125emX}}
\begin{document}

\title{Scaling Effects of Transistor Leakage Current and IR Drop on 1T1R Memory Arrays     \\
}

\author{\IEEEauthorblockN{ Junren Chen, Giacomo Indiveri}
\IEEEauthorblockA{\textit{
Institute of Neuroinformatics},
\textit{University of Zurich and ETH Zurich}, Zurich, Switzerland \\
junren, giacomo@ini.uzh.ch}
}

\maketitle

\newcommand{\MP}[1]{\textcolor{red}{{\bf MP:}~#1}}
\newcommand{\JC}[2]{\textcolor{blue}{{\bf JC:}~#1}}

\begin{abstract}
  1T1R (1-transistor-1-resistor) memory crossbar arrays represent a promising solution for compute-in-memory matrix-vector multiplication accelerators and embedded or storage-class memory.
  However, the size and scaling of these arrays are hindered by critical challenges, such as the IR drop on metal lines and the accumulation of leakage current from the transistors.
  Although the IR drop issue has been extensively investigated, the impact of transistor leakage current has received limited attention.
  In this work, we investigate both issues and highlight how transistor leakage in 1T1R arrays has effects similar to IR drop, which degrades the memory cell sensing margin, especially as the technology node scales down.
  This degradation could pose reliability concerns, particularly where the on/off ratio or sensing margin of memristors is critical.
  We characterized the joint effects of transistor read resistance, transistor leakage current, and IR drop as the array size scales up and the fabrication node scales down.
  Based on a model developed using specifications of a 22\,nm FDSOI technology, we found that an optimal resistance range of memristors exists for good array scaling capability, where the transistor read resistance and the IR drop issue establish a lower resistance boundary, while the transistor leakage issue sets an upper resistance boundary.
  This work provides valuable scaling guidelines for engineering the properties of memristor devices in 1T1R memory arrays.

\end{abstract}


\section{Introduction}
\label{sec:introduction}

Emerging memory technologies, such as resistive random access memory (RRAM), phase-change memory (PCM), and magnetoresistive RAM (MRAM), also known as memristors, are promising candidates for enhancing future memory systems due to their superior characteristics, including fast access speed, high density, and non-volatility~\cite{chen2017emerging}. The memory arrays are commonly organized in a 1T1R crossbar structure and are widely used in hybrid memristive-CMOS processing systems, with transistors serving as selectors for the memristive devices.

These memories are being actively explored to address the memory bottleneck problem in terms of energy and area efficiency.
For example, the success of deep learning has spurred significant interest in developing large 1T1R arrays as compute-in-memory (CIM) accelerators, leveraging their ability to perform highly efficient matrix-vector multiplication (MVM) to meet the growing computational demands and mitigate expensive data movements between memory and processing units for running deep neural networks~\cite{liu2020isscc,wan2022compute,le202364,jung2022crossbar}.
In the context of embedded memory, the demand of the emerging non-volatile memories is driven by Internet-of-Things (IoT) devices beyond 28nm technology nodes, due to the challenges and costs associated with scaling charge-based embedded flash memory~\cite{chih2021design,lee2018embedded,wei2019,chou202022nm,Zuliani2019,Arnaud2020}.
Additionally, 1T1R memory arrays have high potential as memristive crossbar routers (i.e., non-volatile switch matrices) for transmitting voltage pulses among neural computing cores in multi-core spiking neural network (SNN) chips~\cite{dalgaty2024mosaic,2023chen,chen2022reliability}. This approach offers an alternative to CMOS and SRAM-based routers used in these chips~\cite{su2023,su2024}, eliminating static power consumption.



\begin{figure}
\centerline{\includegraphics[scale=0.23, trim={0cm 0cm 0cm 0cm},clip]{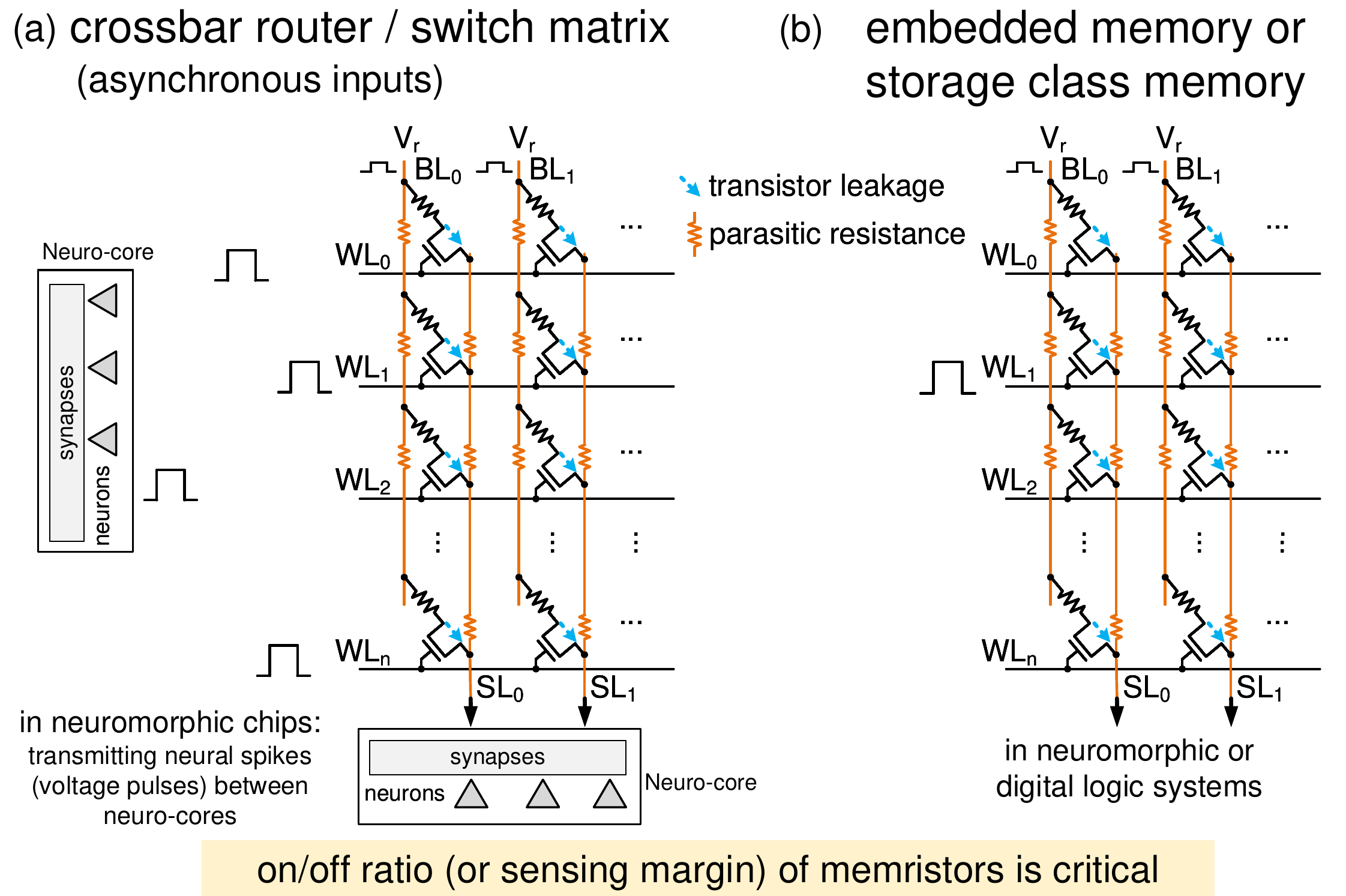}}
\caption{Illustration of non-ideal factors, including metal line resistance and transistor leakage current, in 1T1R arrays for various applications where sensing margin is critical: (a) Preventing signal transmission errors in memristor-based crossbar routers (i.e., switch matrices), which are used to transmit signals between neural computing cores in multi-core spiking neural network (SNN) neuromorphic chips~\cite{dalgaty2024mosaic,2023chen,chen2022reliability}.
(b) Avoiding bit flips during memory cell sensing in embedded memory or storage class memory (SCM) designs. The IR drop issue has been extensively studied while the the transistor leakage current gained limited attention.
}
\label{fig:Intro}
\end{figure}

However, scaling these memory crossbar arrays poses significant physical challenges. While the impact of IR drop on metal lines due to parasitic resistance has been extensively investigated~\cite{2023chen,qin2022hybrid,liao2020compact,yu2015scaling,liang2013effect}, the effect of transistor leakage current has not received comparable attention.
As CMOS technology continues to scale, increased transistor leakage is becoming a prominent challenge~\cite{2003LeakPower,kim2010challenges}.
The reduced memory cell sensing margin due to the combined effects of IR drop and transistor leakage current can significantly degrade the reliability and performance of 1T1R memory arrays, particularly in scaled technology nodes. This issue is especially critical in applications where the on/off ratio or sensing margin is a key performance metric (in Fig.~\ref{fig:Intro}).

To dive into these joint scaling effects, we developed a comprehensive model that unifies the influence of metal line resistance, transistor read resistance and transistor leakage current on the sensing margin of memory cells in the array.
Based on this model, we characterized their joint effects as the array size scales up and the fabrication node scales down. Furthermore, we validated the mathematical model through circuit simulations using a 22nm FDSOI technology node. The major contributions of this work are:
\begin{enumerate}
    \item Identification of the scaling effects of transistor leakage currents on degrading memory cell sensing margin in 1T1R arrays, which was previously overlooked.
    \item An easy-to-use mathematical model that can accurately capture the effects of both IR drop on metal lines and transistor leakage currents for analyzing the real sensing margin (or on/off ratio) of memristors to the sensing circuits for a given array size and technology node. Using parameters such as metal line sheet resistance and geometry (length and width), transistor leakage current, and memristor resistance, the model enables accurate sensing margin predictions without needing SPICE simulations of large memory arrays.
\end{enumerate}

In the next Section we present our methodology and model; in Section~\ref{sec:res} we report the results of our analysis, and finally in the last Section we present concluding remarks on our findings.





\section{Methods}
\label{sec:meth}
\subsection{1T1R Column Modeling When Sensing the Cells}

\begin{figure}
\centerline{\includegraphics[scale=0.3, trim={0.5cm 0cm 0cm 0cm},clip]{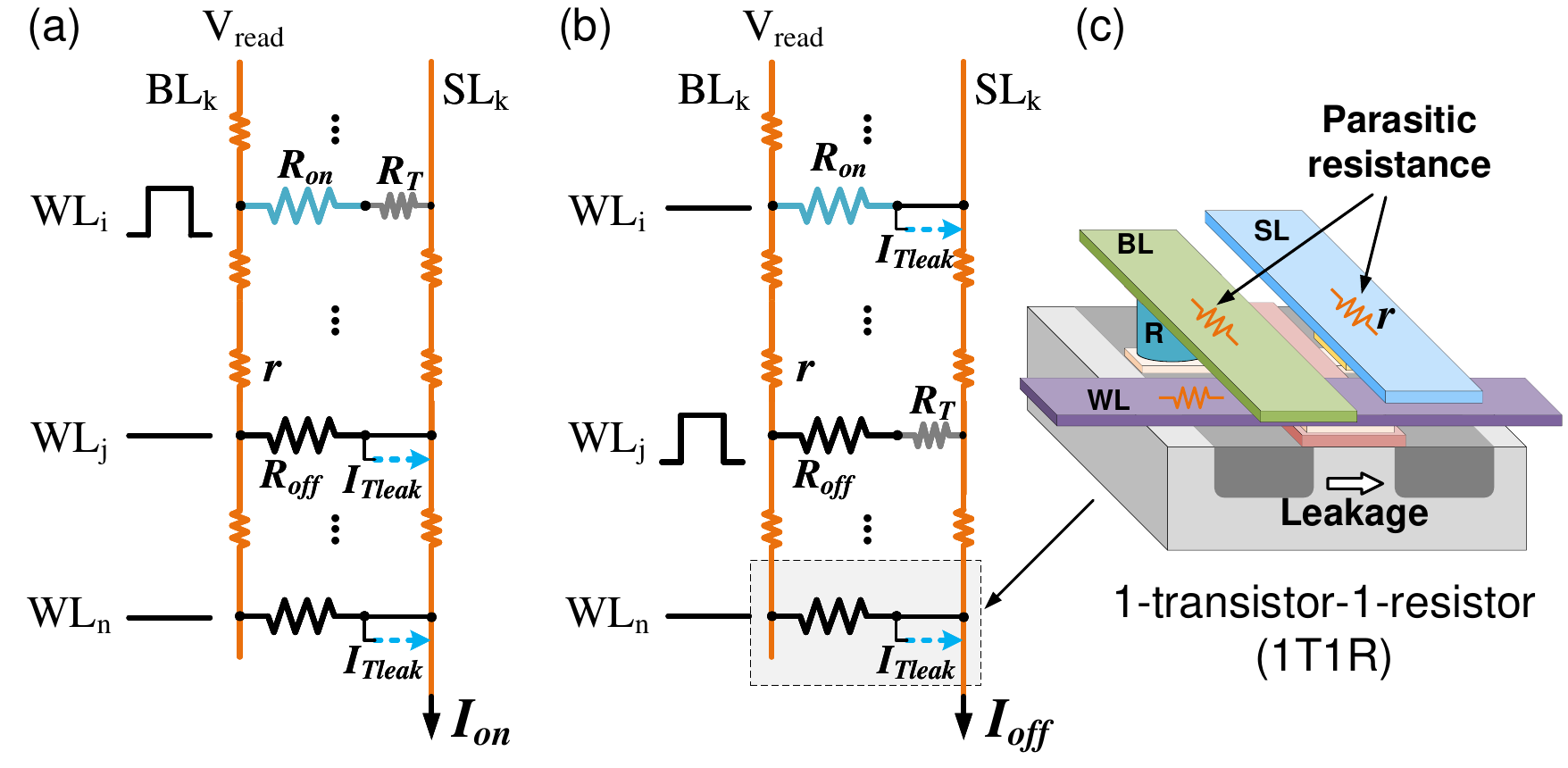}}
\caption{The model of one column in 1T1R crossbar array when sensing a memory cell including parasitic metal line resistance and transistor leakage current. Assuming $n > j > i$. Each transistor which is turned off contributes $I_{Tleak}$, while which is turned on contributes $R_T$ in serial with the resistance of the memristor.
(a) The overall read current $I_{on}$ when sensing $R_{on}$.
(b) The overall read current $I_{off}$ when sensing $R_{off}$.
(c) The physical view of 1T1R cell that depicts the metal line resistance and the transistor leakage current.
}
\label{fig:1t1r_model}
\end{figure}

In the ideal case without IR drop and transistor leakage currents, the on/off ratio of fabricated memory cells can be defined as
\begin{equation}  \label{eq:k}
    k = \frac{V_{read}}{R_{on}}/\frac{V_{read}}{R_{off}} = \frac{R_{off}}{R_{on}}
\end{equation}

where $k$ stands for the on/off ratio of the fabricated memristors. $V_{read}$ is the read voltage across the memristors. $R_{on}$ is the resistance of the ``on" state, while $R_{off}$ is the resistance of the ``off" state.

In practice, however, when the non-ideal factors come into play, as shown in Fig.~\ref{fig:1t1r_model}, the on/off ratio becomes
\begin{equation} \label{eq:k_p}
    k' = I_{on}/I_{off} = \frac{\frac{V_{read}}{R_{on}+R_T+nr} + (n-1)I_{Tleak}} {\frac{V_{read}}{kR_{on}+R_T+nr} + (n-1)I_{Tleak}}
\end{equation}

Here, $k'$ stands for the real on/off ratio of the memristors to the sensing circuit at the end of the metal lines. $R_T$ is the read resistance of a transistor when the gate is turned on, and $I_{Tleak}$ is the leakage current flowing through the transistor when the gate is off. $r$ is the unit line resistance between two adjacent memory cells, which contributes to the IR drop on BLs and SLs. $n$ is the number of cells in one dimension (column) of the array, representing the array size. Since WLs connect to transistor gates and the currents on WLs during reads are negligible, IR drop on WLs is disregarded. It is evident that while the influence of $R_T$ is independent of array size, the metal line IR drop and the cumulative transistor leakage are correlated with array size.
It should be noted hereafter that $k'/k$ is the normalized sensing margin, quantifying how much of the original sensing margin remains in the presence of non-idealities.
The effect of $r$ has been extensively studied in the literature. Due to parasitic resistance on metal lines, the read voltage applied to memory cells down the line is lower than intended, leading to reduced output currents. This degradation can affect the reliability of memory devices, as well as the accuracy of computations in analog CIM accelerators. However, the scaling effects of transistor leakage in the 1T1R array have received limited attention, where $I_{Tleak}$ in Eq.~\ref{eq:k_p} was lacking and not addressed in previous works. This factor will play an important role that degrades the sensing margin as the array size scales up and the technology node scales down (in Section~\ref{sec:res}).

\begin{figure*}[t]
\centerline{\includegraphics[scale=0.42]{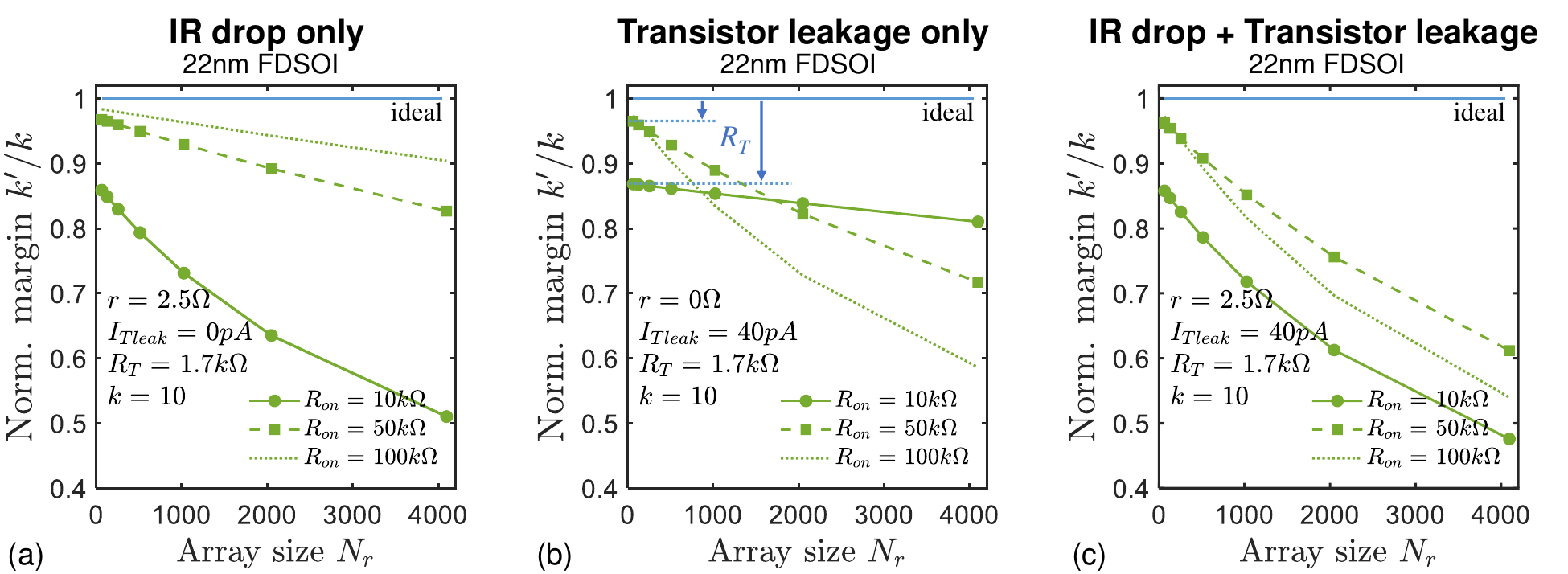}}
\caption{Impact of metal line IR drop and transistor leakage currents on the effective sensing margin of memory cells as the array size increases from 64 to 4096 ($N_r$ denotes $n$ in Eq.~\ref{eq:k_p}). 
(a) IR drop degrades the effective on/off ratio (sensing margin), with $V_{read}$ is 0.2V. Higher resistance values ($R_{on}$) mitigate the effect of IR drop.
(b) Transistor leakage in 1T1R arrays similarly degrades the sensing margin. In contrast to IR drop, higher resistance boosts the effect of transistor leakage and worsens the sensing margin. The influence of $R_T$ is also annotated.
(c) The combined scaling effects of IR drop and transistor leakage. When both factors are considered, the optimal resistance for size scaling shifts, highlighting potential trade-offs are involved in array size scaling.
}
\label{fig:joint_effects}
\end{figure*}

\subsection{1T1R Array Circuit Simulations}
The read resistance of the transistor is extracted by simulating a transistor while reading the 1T1R cell, as described in~\cite{2023chen}. The transistor size is selected to ensure appropriate driving capacity for programming memristors ($>150uA$, $210nm/40nm$). Regular threshold voltage transistors are used instead of high-threshold ones which typically have lower leakage currents in the same technology node. This decision is made for two main reasons. First, we aim to explore and analyze the effects of leakage currents, rather than optimizing for minimal leakage. Second, high-threshold transistors have lower speed and require a larger area to provide adequate drive strength, which in turn decreases both the performance and density of the memory array. This is also why high-threshold transistors are usually not used as selectors in 1T1R arrays when the impact of leakage is not a major concern. Based on simulations and PDK, we get $I_{Tleak} = 40pA$ (at $V_{read}=0.2V$) and estimate $R_T = 1.7\,k\Omega$, $r \approx 2.5\,\Omega$ in 22nm node.
Furthermore, to validate the proposed mathematical model, the schematics of 1T1R array were constructed with varying array sizes and simulated using the Cadence Spectre Simulator in a 22nm FDSOI technology.

\section{Results}
\label{sec:res}

\subsection{Joint Scaling Effects of Transistor Leakage and IR Drop}
\label{sec:joint}


Using the model of Eq.~\ref{eq:k_p} which unifies the non-idealities of both metal line resistance and transistor leakage, their individual and joint impacts on the sensing margin of memory cells can be studied.
If we focus solely on IR drop, we may draw the incomplete conclusion that a higher memristor resistance always improves array scalability. For instance, when only IR drop is considered in Fig.~\ref{fig:joint_effects} (a), a higher resistance ($R_{on}=100k\Omega$) appears favorable for reducing the degradation of sensing margin. However, the accumulation of transistor leakage currents also plays a critical role, particularly as the array size scales up, with effects similar to IR drop. As shown in Fig.~\ref{fig:joint_effects} (b), this leakage degradation becomes more pronounced at higher memristor resistance values, where a lower resistance ($R_{on}=10k\Omega$) is more effective in mitigating the impact of $I_{Tleak}$ on sensing margin. 
Interestingly, Fig.~\ref{fig:joint_effects} (c) shows that when both IR drop and transistor leakage currents are taken into account, an intermediate resistance ($R_{on}=50k\Omega$) becomes optimal, which is counter-intuitive.
This observation prompts further investigation into whether an optimal resistance range for memristors exists — one that maximizes scalability while balancing the impacts of IR drop and transistor leakage.

\subsection{The Best Resistance Range for Array Scaling}
\label{sec:best}

\begin{figure}[htbp]
\centerline{\includegraphics[scale=0.35, trim={0 0.1cm 0 0},clip]{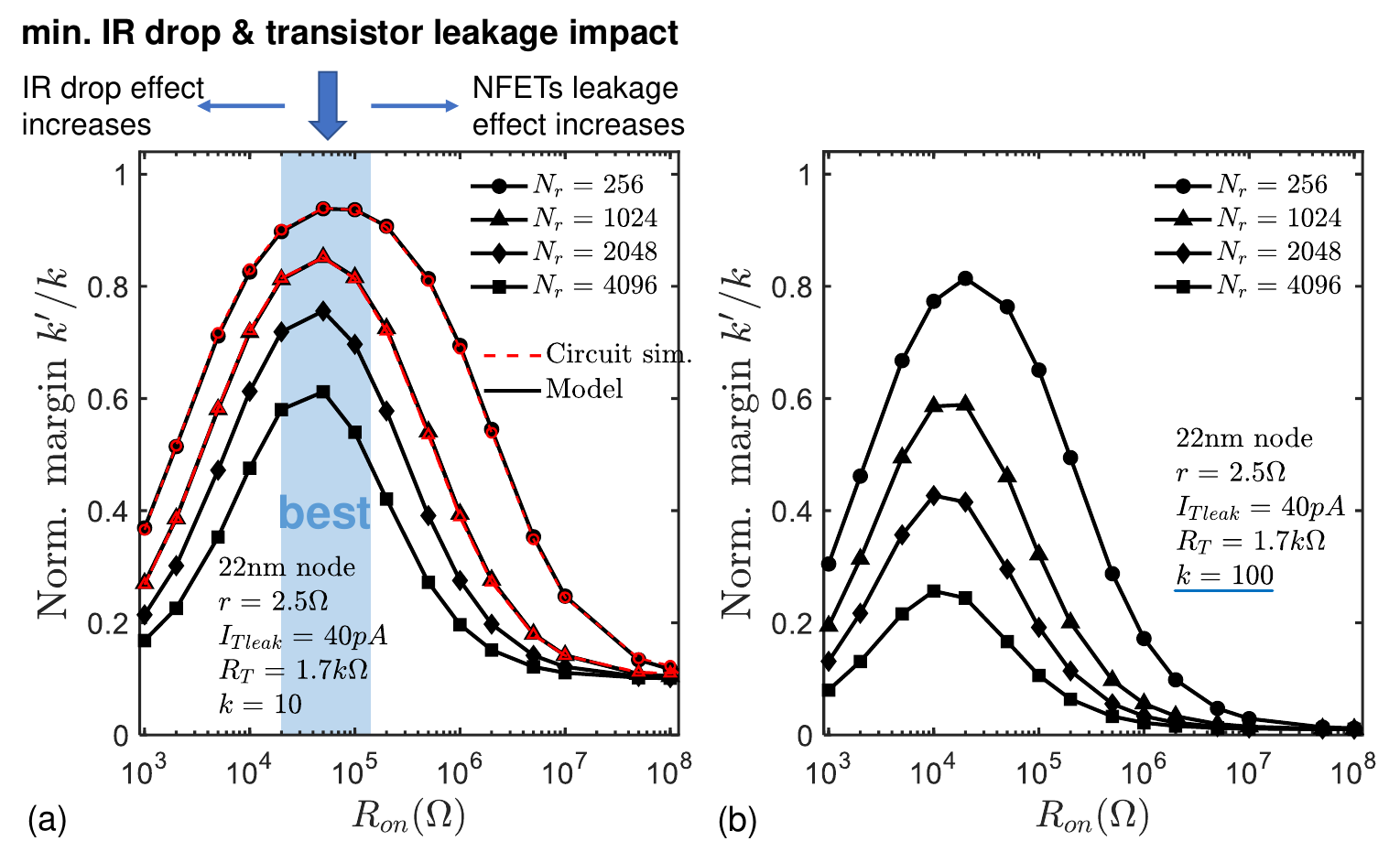}}
\caption{(a) For optimal array size scaling while maintaining a high sensing margin, $20k\Omega<R_{on}<150 k\Omega$ is ideal (at $k = 10$). The mathematical model matches the circuit simulations very well. Each red data point corresponds to a circuit simulation. (b) The size scaling effect at $k=100$.
}
\vspace{-0.2cm}
\label{fig:best_R}
\end{figure}

We explore the change of sensing margin by sweeping $R_{on}$ from $10k\Omega$ to $100 M\Omega$ while adjusting the array size from 256 to 4096, under the condition $k = 10$ (which implies $R_{off}$ ranges from $100k\Omega$ to $1G\Omega$), as depicted in Fig.~\ref{fig:best_R} (a). The red dashed lines represent the circuit simulation results, which is a validation that the model matches the circuit simulations very well. This agreement allows us to use the model for further exploration of various parameters without the need for extensive SPICE simulations of heavy memory arrays.
It also shows that the metal line IR drop effect begins to emerge if the resistance of memory cell decreases to $<20 k\Omega$, while the transistor leakage current accumulation starts to dominate and becomes the primary issue when $R_{on}>$ a few hundred of $k\Omega$. The best $R_{on}$ range for scaling up the array size is $20k\Omega$ to $150 k\Omega$ in 22nm node. 
If a tenfold increase in sensing margin is desired, Fig.~\ref{fig:best_R} (b) indicates that the best resistance range shifts slightly lower, and the non-idealities have a greater impact on the percentage drop in effective sensing margin.

\begin{figure}[htbp]
\centerline{\includegraphics[scale=0.4, trim={0 0.1cm 0 0cm},clip]    {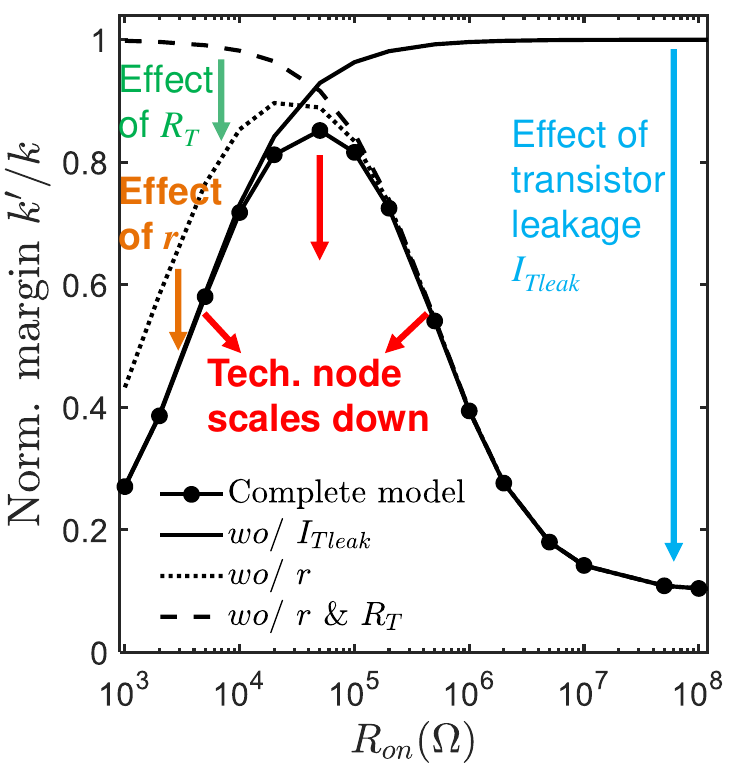}}
\caption{
Detailed scaling effects illustration of $R_T$, $r$ and $I_{Tleak}$, using the model of $N_r=1024$ in Fig.~\ref{fig:best_R} (a) in 22nm node.
As the CMOS technology node scales down, IR drop and transistor leakage will influence more. The scaling trend of sensing margin degradation as technology node scales down will be similar to that of size scaling up.
}
\vspace{-0.2cm}
\label{fig:scaling_details}
\end{figure}

To understand the individual impact of the physical non-idealities, we take the model of $N_r=1024$ in Fig.~\ref{fig:best_R} (a), and remove the individual parameters one by one to visualize the changes. Figure~\ref{fig:scaling_details} highlights the three key factors affecting the sensing margin: $R_T$, $r$ and $I_{Tleak}$. It illustrates that an optimal resistance range of memoristors exists for scaling 1T1R memory arrays. The lower resistance boundary is determined by the read resistance of transistors and the IR drop issue, while the upper resistance boundary is set by the transistor leakage issue. Compared to ideal values, IR drop reduces $I_{on}$, whereas transistor leakage increases $I_{off}$. It can be interpreted as an increase in $R_{on}$ and a decrease in $R_{off}$, respectively.
This provides useful guidelines for memristive device engineering.
For example, to support the implementation of an array with size 512, where the effect of $r$ and $R_T$ are comparable ($1.28 k\Omega$ vs. $1.7 k\Omega$), $R_{on}$ should be fabricated to $>20 k\Omega$ to gain a sensing margin around 85\%. However, $R_{on}$ should not exceed the $M\Omega$ range, as the overall
transistor leakage current will overwhelm the current from the
memory cells in their off state.


Furthermore, $I_{Tleak}$ increases as the technology node scales down~\cite{2003LeakPower,datta2022toward}, and $r$ also increases due to the shrinking geometry of metal interconnects. These combined effects pose challenges for size scaling of 1T1R arrays. Notably, the effects of scaling down of the technology node for a fixed array size can be likened to the scaling up of the array size for a given technology node.

It is worth noting that in this work, we have demonstrated and validated the model using a 22nm FDSOI technology as a case study. However, the model is expected to be general to other technology nodes and can efficiently estimate the effects of various parameters based on the PDK.

\subsection{Trade-off: Sensing Margin vs. Power Consumption \& Reliability}
\label{sec:trade-off}

\begin{figure}[htbp]
\centerline{\includegraphics[scale=0.4, trim={0 0.1cm 0 0cm},clip]    {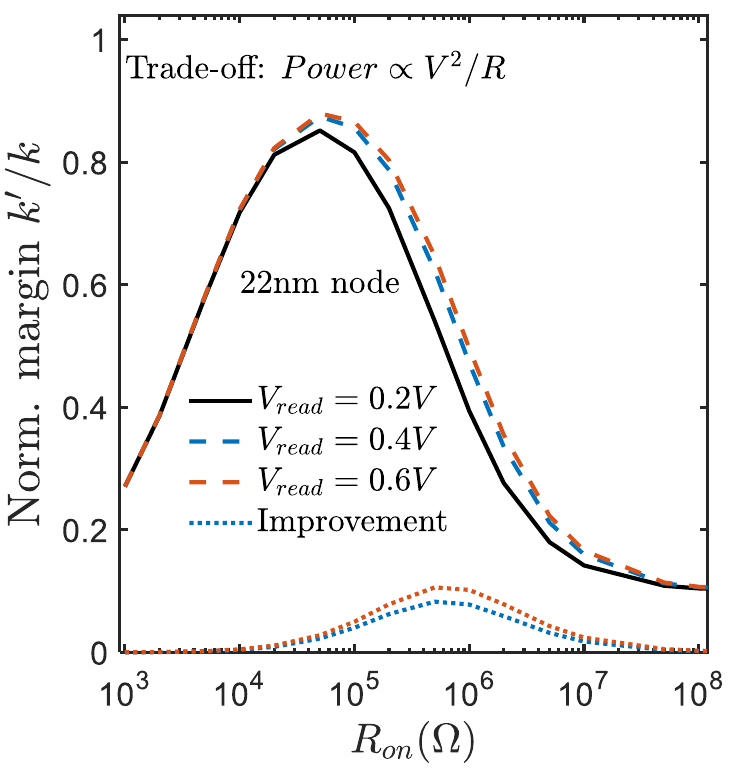}}
\caption{
The effect of read voltage. The blue dotted line (subtracting the black line from the blue dashed line) shows the improvement of sensing margin by increasing $V_{read}$ from 0.2V to 0.4V, at the cost of increased power consumption. The orange dotted line shows the improvement at $V_{read}=$ 0.6V.
}
\vspace{-0.2cm}
\label{fig:Vread}
\end{figure}

Figure~\ref{fig:Vread} shows the influence of $V_{read}$. Increasing the read voltage from 0.2V to 0.4V and 0.6V can compensate for the sensing margin drop by max. $8\%$ and $11\%$, respectively, though this results in a quadratic increase in read power consumption.  Correspondingly, $I_{Tleak}$ increases from $40pA$ to $55pA$ and $74pA$ across this voltage range, indicating that the benefits of further increasing $V_{read}$ diminish as the voltage rises. Thus, using high-threshold transistors would be more advantageous than continuously increasing $V_{read}$ at some point, if the impact of leakage is critical. 
It should be noted that Fig.~\ref{fig:Vread} is obtained under the assumption that the I-V relationship of the memory cell is approximately linear in this voltage range, adhering to Ohm's Law. In the case that the memory cell current increases exponentially with increased read voltage ($I \propto e^V$), the dashed lines would increase with higher improvements. Meanwhile, read power consumption would be $\propto Ve^V$. 
Moreover, this compensation strategy is effective only within the resistance range where transistor leakage current is significant, and not where the metal line IR drop effect predominates. In addition, a trade-off exists between the sensing margin and the reliability of reading memory cells: while compensating for the sensing margin reduction caused by transistor leakage, the read voltage should not be excessively high to mitigate the risk of faulty write operation.

\section*{Conclusion}
We presented a model and corresponding analysis that produced counter-intuitive results, which can be useful in guiding the design of scaled 1T1R memory arrays.
For the case study in 22nm technology node, our findings show that the optimal resistance range for memristors that ensures effective scaling of 1T1R arrays is $20 k\Omega<R_{on}<150 k\Omega$ when the on/off ratio is around 10.
The lower resistance boundary is determined by the read resistance of the transistor and the metal line IR drop issue, while the upper resistance boundary is set by the transistor leakage issue. Technology node scaling down hinders array size scaling up due to the sensing margin degradation of memory cells, resulting from the increase of parasitic metal line resistance and transistor leakage currents. Increasing read voltage could compensate for the sensing margin loss, at the expense of a dramatic increase in power consumption and a decrease of reliability.

\section*{Acknowledgment}

This work was supported by the Innovation Program for ECSEL grants ANDANTE (grant agreement No. 876925).

\section*{Code availability}
The open source code for the modeling and analysis can be found on GitHub (\href{https://github.com/Junren-Chen/ISCAS_2025.git}{https://github.com/Junren-Chen/ISCAS\_2025.git}).

{
\bibliographystyle{IEEEtran}
\bibliography{reference}
}

\end{document}